%% file: main.tex
\documentclass[a4paper,UKenglish]{lipics}
\usepackage{microtype}%if unwanted, comment out or use option "draft"

\usepackage{tikz,pgfplots}
\usetikzlibrary{matrix}
\usepgfplotslibrary{groupplots}
\pgfplotsset{compat=newest}

\usepackage{wrapfig}

\usepackage{comment}

\newcommand{\remove}[1]{}

\title{
Flat Parallelization}
\author[1]{Vitaly Aksenov}
\author[2]{Petr Kuznetsov}
\affil[1]{INRIA Paris, France and ITMO University, Russia\\
  \texttt{aksenov.vitaly@gmail.com}}
\affil[2]{LTCI, T\'el\'ecom ParisTech, Universit\'e Paris-Saclay\\
  \texttt{petr.kuznetsov@telecom-paristech.fr}}
\authorrunning{V. Aksenov and P. Kuznetsov} %mandatory. First: Use abbreviated first/middle names. Second (only in severe cases): Use first author plus 'et. al.'

\Copyright{Vitaly Aksenov and Petr Kuznetsov}%mandatory, please use full first names. LIPIcs license is "CC-BY";  http://creativecommons.org/licenses/by/3.0/

\remove{
\serieslogo{}%please provide filename (without suffix)
\volumeinfo%(easychair interface)
  {Billy Editor and Bill Editors}% editors
  {2}% number of editors: 1, 2, ....
  {Conference title on which this volume is based on}% event
  {1}% volume
  {1}% issue
  {1}% starting page number
}

\begin{document}
\maketitle

\input{abstract.tex}

\input{introduction.tex}

\input{algorithm.tex}
\input{experiments.tex}
\input{conclusion.tex}

\bibliography{references}

\end{document}

%% file: abstract.tex
\begin{abstract}                                           

There are two intertwined factors that affect performance of concurrent
data structures: the ability of processes to access the
data in parallel and the cost of synchronization.
It has been observed that for
a large class of ``concurrency-unfriendly'' data structures,
fine-grained parallelization does not pay off: an implementation based
on a single global lock outperforms fine-grained solutions.   
The \emph{flat combining} paradigm exploits this by ensuring that
a thread holding the global lock sequentially \emph{combines} requests and then 
executes the combined requests on behalf of concurrent threads.

In this paper, we propose a synchronization technique that unites
flat combining and parallel bulk updates borrowed from parallel algorithms
designed for the PRAM model.
The idea is that the combiner thread assigns waiting threads to
perform concurrent requests in parallel.

We foresee the technique to help in implementing efficient ``concurrency-ambivalent'' data
structures, which can benefit from both parallelism and serialization,
depending on the operational context.      
To validate the idea, we considered \remove{\emph{heap-based}} implementations of
a \emph{priority queue}.
These data structures exhibit two important features:
concurrent remove operations are likely to conflict and thus may
benefit from combining,
while concurrent insert operations can often be at least partly applied in parallel thus may benefit from parallel batching.
We show that the resulting \emph{flat parallelization} algorithm performs well compared to state-of-the-art
%skiplist-based and flat-combining
priority queue implementations.

\end{abstract}

%% file: introduction.tex
\section{Introduction}

There are two intertwined factors that affect performance of concurrent
data structures: the ability of threads to access the
data in parallel and the cost of synchronization.
For many data structures, such as \textsf{linked lists},
\textsf{skiplists}, \textsf{trees}, etc.,
a high degree of parallelism can be achieved by fine-grained
synchronization among concurrent threads.
%which results them to
%outperform simple coarse-grained implementations based on a single
%global lock.   
%
It has been observed, however, that for
a large class of ``concurrency-unfriendly'' data structures, such as 
\textsf{queues}, \textsf{stacks}, and \textsf{priority queues}, 
fine-grained parallelization does not pay off.
%an implementation based
%on a single global lock outperforms fine-grained solutions.
Operations on such data structures are likely to conflict on data
``hotspots'' and, as a result, the costs  incurred by fine-grained synchronization mechanisms
(low-level locks or lock-free synchronization primitives) do not allow
us to overcome the performance of simple coarse-grained solutions based on a single
global lock.

%Inspired by this observation,
In the \emph{flat combining} paradigm~\cite{hendler2010flat}, 
a thread holding the global lock sequentially
\emph{combines} requests published by concurrent threads and then 
executes the combined requests on their behalf.
Besides providing \emph{starvation-freedom} (every thread makes
progress), the approach enables an efficient execution of combined
updates: e.g., a bunch of updates can often be performed by the
combiner thread in a single
pass over the data structure. Also, some updates, e.g., concurrent
operations \texttt{pop} and \texttt{push} on a \textsf{stack}, can \emph{eliminate} each
other without touching the data structure.        

However, while the combiner is busy performing the requests, other
threads remain idle and the available multi-processor
computational power remains unused.
In this paper, we propose a technique that leverages this power,
while preserving the advantages of flat combining.
In particular, we consider the \emph{bulk-update}
technique~\cite{paul1983parallel} devised for the PRAM model of
synchronous parallel computing.
At a high level, the combiner (or the \emph{leader} thread in our
terminology) distributes ``parallelizable'' updates in a bulk among
the waiting (\emph{worker}) threads, so that they could perform the updates in
a conflict-free way.       

We foresee the technique to help in implementing efficient ``concurrency-ambivalent'' data
structures, which can benefit from both parallelism and serialization,
depending on the operational context.      
To validate the approach, we considered \emph{binary heap-based} implementations of
a \textsf{priority queue}.
This data structure exhibits two important features:
concurrent \texttt{extractMin} operations are likely to conflict and thus may
benefit from combining,
while concurrent \texttt{insert} operations can often be (at least
partially) applied in parallel and thus may benefit from parallel batching.
We show that the resulting \emph{flat-parallelization} algorithm performs well compared to state-of-the-art
%skiplist-based and flat-combining
concurrent priority queue implementations.

%% file: algorithm.tex
\section{Algorithm}                                            
As a basis of our algorithm we took the binary heap based implementation of the priority queue. Despite
the fact that the binary heap algorithm has one of the worst asymptotics compared to other heaps,
it is easy to implement and its array-based implementation
\cite{thomas2009priorityqueues} is fast because it does not induce overhead on memory management.

We briefly describe the sequential array-based binary heap implementation on which we based 
our parallel bulk-update algorithm. Note, that the implementation
of \texttt{insert} operation differs from the one described in \cite{thomas2009priorityqueues}.
The heap is represented as a 1-indexed array
$a$ with size $S$ where node $v$ has children $2v$ and $2v + 1$. The heap should satisfy the property
that for any node $v$ the value in $v$ is less than the values in the children. The two operations
\texttt{extractMin} and \texttt{insert} are performed as follows:
\begin{itemize}
\item \texttt{ExtractMin} operation swaps $a[size]$ with $a[1]$ and
  then performs a procedure \texttt{sift down} to restore the heap property.
  We start the procedure from the root. At each iteration, we are located in some node $v$.
  We compare $a[v]$ with the values in children and consider two cases. $a[v]$ is less than the values
  in the children then the property of the heap is satisfied and
  we stop our operation. Otherwise, we choose the child $c$, either $2v$ or $2v + 1$, with the smallest value, swap values $a[v]$
  and $a[c]$, and continue with $c$.
\item During \texttt{insert(x)} operation we consider the path from the root to node $S + 1$.
  The newly inserted value $x$ should be somewhere on that path, thus we only need to insert it in the proper place
  and shift the rest of the path one level down. We initialize a variable $val = x$ and start traversing
  the path from the root to $S + 1$. Suppose, that we are currently located in node $v$. We compare $val$ and $a[v]$ and
  consider two cases. If $val$ is less than $a[v]$ then we swap these values and continue with the corresponding children $c$,
  i.e., the next node on the path. Otherwise, $val$ is bigger than $a[v]$ and we simply continue with $c$.

  The complexity of classical \cite{thomas2009priorityqueues} and described implementations are $O(\log n)$. Nevertheless,
  the original apprroach works faster on average, because it generally does not need to traverse
  the whole height of the heap.
\end{itemize}

Now, we are ready to describe our parallel bulk-update algorithm.
Our algorithm separates \texttt{extractMin}
and \texttt{insert} and processes, firstly, a batch of \texttt{extractMin} requests and then a batch of
\texttt{insert} requests. 
In a few words, 
for $k$ \texttt{extractMin} operations we swap $k$ smallest values
with $a[S - k + 1], a[S - k + 2], \ldots, a[S]$ and perform
$k$ parallel \texttt{sift down} procedures, each in a separate thread. The algorithm benefits from the fact
that sift downs are in general operate on different subtrees.
Note, that for this algorithm we have to find $k$ smallest values first.
For $k$ \texttt{insert} operations one of the threads
starts inserting all the values simultaneously and traverse the heap towards nodes $S + 1, \ldots, S + k$.
Sometimes, it splits its set of values and gives the tasks to other threads.
Now, we describe each bulk-udpate operation in more details.
\begin{itemize}
\item $k$ \texttt{extractMin} operations. As discussed earlier we are provided with the nodes with the smallest $k$
values. At first, we set the special field $locked$ in each of these nodes.
Then, we swap these values with the $k$ latest values in the heap, i.e., $a[S - k + 1], \ldots, a[S]$.
The nodes where were the smallest values now become the start positions of $k$ parallel
sift down procedures, each performed by a separate thread. The $locked$ field
specifies if the value in the node is currently under swap. Each thread starts its own sift down from the
corresponding node. In each iteration it considers some node $v$. The thread waits while the children of 
$v$ has $locked$ field set. When all the children becomes unlocked, the thread compares the value $a[v]$
with the values in the children. If $a[v]$ is smaller then we unset the $locked$ field of $v$ and stop the
procedure. Otherwise, we choose the child $c$ with the smallest value, we set $locked$ field of $c$, swap
$a[v]$ with $a[c]$, unset $locked$ field of $v$ and continue the next iteration with $c$.
\item $k$ \texttt{insert} operations. As in the sequential implementation of \texttt{insert}, $k$ newly inserted
values should be on the paths from the root to nodes $S + 1, \ldots, S + k$. Primarily, we set
the special field $split$ in the nodes that have some of the nodes $S + 1, \ldots, S + k$ in both subtrees.
Then one thread starts from the root while other threads wait on the nodes with $split$ set. 
Note, that there are exactly $k - 1$ $split$ nodes.
The thread that starts from the root is provided with the set of $k$ inserted values. Each thread
on each iteration is located at some node $v$. If the smallest value from its set is less than $a[v]$, it 
puts the smallest value from the set to $a[v]$ while inserting $a[v]$ into set, otherwise, the thread
does nothing.
After, it checks whether the $split$ field is set in node $v$.
If it is not set,
then the thread continues with the corresponding child. Otherwise, the thread splits the set into two parts,
in proportion of how much nodes from $S + 1, \ldots, S + k$ are in the left subtree and in the right subtree.
The thread gives the set for the right subtree to the thread that waits at the node $v$ and unsets $split$.
By itself, the thread continues with the left children, while the waken up thread starts with the right child.
\end{itemize}

We argue, that the described algorithm, given the preliminary work done, performs a bulk-update on $k$ operations in
$O(k \log n)$ work and $O(k + \log n)$ span. Thus, providing us with linear speedup when $k \leq \log n$.

Now, given the parallel batching algorithm we explain how we incorporate it together with the flat combining
technique, probably omitting some optimizations.
\begin{itemize}
\item A thread puts a request of the operation in the queue. Each request consists of the type of the operations,
either \texttt{extractMin} or \texttt{insert}, and the status of the request. The status could be one of the three
types: \texttt{PUSHED}, \texttt{SIFT} and \texttt{FINISHED}. The freshly pushed request has \texttt{PUSHED} state.
\item The thread tries to take a lock on the data structure. If it succeeds then it becomes a leader.
\item If the thread becomes a leader it performs the following:
  \begin{itemize}
    \item The leader takes all non-performed operations from a queue. Sorts them by the type and the value. Let $E$ be the number
      of \texttt{extractMin} requests and $I$ be the number of \texttt{insert} requests.
    \item The leader finds $E$ nodes with the smallest values and sets their $locked$ fields.
      Then he swaps their values with the $\min(E, I)$ newly inserted values and the rest with
      the latest $E - \min(E, I)$ values in the heap. That is how we combine two types of requests.
      And finally, he sets the state of \texttt{extractMin} requests to \texttt{SIFT} and 
      the state of first $\min(E, I)$ \texttt{insert} requests to \texttt{FINISHED}.
    \item The leader, probably, performs \texttt{extractMin} operation by itself and then waits while the workers
      with \texttt{extractMin} request set their status to \texttt{FINISHED}.
    \item The leader sets the $split$ field of the nodes that have nodes $S + 1, \ldots, S + (I - \min(E, I))$
      in both subtrees. Then it sets the state of remaining \texttt{insert} requests to \texttt{SIFT}.
    \item The leader, probably, performs \texttt{insert} operation by itself and then waits while the workers
      with \texttt{insert} request set their status to \texttt{FINISHED}.
  \end{itemize}
\item If the thread is not a leader, then it spins until the status of its request becomes \texttt{SIFT}.
Then it performs its operation and sets the status to \texttt{FINISHED}.
\end{itemize}

One could calculate the number of remote menory accesses during one operation. It does not exceed $O(P + \log S)$,
where $P$ is the number of threads working on the priority queue and $S$ is the current size of the heap.

%% file: experiments.tex
\section{Experiments}

\begin{figure}
\centering
\input{plot.tex}
\captionsetup{justification=centering}
\caption{Experimental results}
\label{fig:experiments}
\end{figure}

For our experiments, we used 4-processor AMD Opteron 6378 2.4 GHz server with
16 threads per processor (yielding 64 threads in total), 512 Gb of RAM, running Ubuntu 14.04.5.
It has Java 1.8.0\_111-b14 and HotSpot JVM 25.111-b14.

We compare our algorithm (FC Parallel) against five implementations: flat combining with binary heap (FC Binary \cite{hendler2010flat}),
flat combining with pairing heap (FC Pairing \cite{hendler2010flat}), lazy lock-based skip-list (Lazy SL),
lock-free skip-list similar to implementation in java.concurrency package (Lock-free SL) and
and coarse-grained binary heap (Coarse Binary).
We are aware of Linden-Johnson algorithm, but we do not have its Java implementation.
For more information
about concurrent priority queue implementations we refer the reader to survey
\cite{gruber2015practical}. The code is available at \url{https://github.com/Aksenov239/FC-heap}.

We provide results for four different settings: the queue is prepopulated with $8 \cdot 10^5$ or $8 \cdot 10^6$ random
integer values, and the inserted values are from $[0, 1000]$ or $[0, 2^{31} - 1]$.
For the depicted plots~\ref{fig:experiments}, we assumed a workload with 50\% \texttt{extractMin} and 50\% \texttt{insert}
operations. Each point is averaged over 5 runs of 10 seconds with the warmup of 10 seconds.

Our algorithm performs badly on the small number of processors for two reasons.
First, the algorithm induces overhead on the preparation and combining parts of flat combining.
Second, our insert operation incurs overhead with respect to
the state-of-the-art algorithm \cite{thomas2009priorityqueues}.
But when the number of processors increases, the parallelization start overwhelming these disadvantages.
Starting from 20 threads our algorithm outperforms all other algorithms.
% except for coarse-grained.
%We could suppose that on higher number of threads the throughput of our algorithm
%could be significantly increased further.

At the same time, the results
of coarse-grained algorithm are suprisingly good in compare to flat combining approaches.
But by common sense, flat combining implementations should perform better.
We link this to the fact that the coarse-grained algorithm is implemented using 
the reentrant lock from Java's concurrent package while our flat combining implementation were
written from scratch identically to the original implementation in C++.

There two reasons why our algorithm slower than the coarse-grained on almost all settings.
First, with the increase of the intial size the ratio of the time spent on the preparation work of the
leader and the time spent during the parallel algorithm decreases. That is why the gap between algorithms shrinks when size increases.
Second, the insert algorithm of the classical implementation of the binary heap spent less time on average on smaller range of values rather than
on bigger range, while the execution of our algorithm does not change.
Thus, when the range increases the gap between algorithms decreases.

Anyway, we find this very encouraging that our algorithm on the settings of initial size $8 \cdot 10^6$ with values from $[0, 2^{31} - 1]$ outperforms
the coarse-grained implementation on 63 processors.

%% file: plot.tex
\begin{tikzpicture}
   \begin{groupplot}[
       group style={
           group size= 2 by 2,
       },
       height=5cm,
       width=5cm,
   ]
   \nextgroupplot[ylabel=Size $8\cdot 10^5$, title=Range: $10^4$, cycle list name=color]
       \addplot table {data/comparison_throughput_800000_10000_FCParallelHeapv2.dat};\label{plots:fcparallel}
       \addplot table {data/comparison_throughput_800000_10000_FCBinaryHeap.dat};\label{plots:fcbinary}
       \addplot table {data/comparison_throughput_800000_10000_FCPairingHeap.dat};\label{plots:fcpairing}
       \addplot table {data/comparison_throughput_800000_10000_LazySkipListHeap.dat};\label{plots:lazyskiplist}
       \addplot table {data/comparison_throughput_800000_10000_SkipListHeap.dat};\label{plots:skiplist}
       \addplot table {data/comparison_throughput_800000_10000_BlockingHeap.dat};\label{plots:blocking}
       \coordinate (top) at (rel axis cs:0,1);% coordinate at top of the first plot

   \nextgroupplot[title=Range: $2^{31}$, cycle list name=color]
       \addplot table {data/comparison_throughput_800000_2147483647_FCParallelHeapv2.dat};
       \addplot table {data/comparison_throughput_800000_2147483647_FCBinaryHeap.dat};
       \addplot table {data/comparison_throughput_800000_2147483647_FCPairingHeap.dat};
       \addplot table {data/comparison_throughput_800000_2147483647_LazySkipListHeap.dat};
       \addplot table {data/comparison_throughput_800000_2147483647_SkipListHeap.dat};
       \addplot table {data/comparison_throughput_800000_2147483647_BlockingHeap.dat};

% -- 8000000

   \nextgroupplot[xlabel=Number of threads, ylabel=Size $8 \cdot 10^6$, cycle list name=color]
       \addplot table {data/comparison_throughput_8000000_10000_FCParallelHeapv2.dat};
       \addplot table {data/comparison_throughput_8000000_10000_FCBinaryHeap.dat};
       \addplot table {data/comparison_throughput_8000000_10000_FCPairingHeap.dat};
       \addplot table {data/comparison_throughput_8000000_10000_LazySkipListHeap.dat};
       \addplot table {data/comparison_throughput_8000000_10000_SkipListHeap.dat};
       \addplot table {data/comparison_throughput_8000000_10000_BlockingHeap.dat};

   \nextgroupplot[xlabel=Number of threads, cycle list name=color]
       \addplot table {data/comparison_throughput_8000000_2147483647_FCParallelHeapv2.dat};
       \addplot table {data/comparison_throughput_8000000_2147483647_FCBinaryHeap.dat};
       \addplot table {data/comparison_throughput_8000000_2147483647_FCPairingHeap.dat};
       \addplot table {data/comparison_throughput_8000000_2147483647_LazySkipListHeap.dat};
       \addplot table {data/comparison_throughput_8000000_2147483647_SkipListHeap.dat};
       \addplot table {data/comparison_throughput_8000000_2147483647_BlockingHeap.dat};
       \coordinate (bot) at (rel axis cs:1,0);% coordinate at bottom of the last plot
  \end{groupplot}
  \path (top-|current bounding box.west)--
       node[anchor=south,rotate=90] {Throughput, mops/s}
       (bot-|current bounding box.west);

  %legend
  \path (top|-current bounding box.north)--
        coordinate(legendpos)
       (bot|-current bounding box.north);
  \matrix[
     matrix of nodes,
     anchor=south,
      draw,
      inner sep=0.2em,
  ] at ([yshift=1ex]legendpos) {
     \ref{plots:fcparallel}& FC Parallel&[5pt]
     \ref{plots:fcbinary}& FC Binary \\
     \ref{plots:fcpairing}& FC Pairing &[5pt]
     \ref{plots:lazyskiplist}& Lazy SL &[5pt]
     \ref{plots:skiplist}& Lock-free SL \\
     \ref{plots:blocking}& Coarse Binary\\
  };
\end{tikzpicture}

%% file: conclusion.tex
\section{Conclusion}
In this paper, we studied new design of concurrent data structures
that unites together flat combining and parallel bulk-update.
We applied this approach to the priority queue data structure. The evaluation suggests
that proposed technique could be an adequate design principle since the demonstrated
performance is comparable to the state-of-the-art algorithms.

Besides performance gains obtained by
exploiting idle threads, there might be other benefits of our \emph{flat parallelization} technique.
First, given a parallel bulk-update algorithm the %flat parallelization 
technique may automatically produce an efficient concurrent counterpart.
For example, one can devise a concurrent \textsf{dynamic forest} with
\texttt{insert-} and \texttt{remove-edge} operations, given the
bulk-update algorithm described in~\cite{acar2017treecontraction}.
%while no concurrent counterpart is known.
Second, the technique can be used to maintain certain non-trivial properties of
data structures, e.g.,  the \emph{strict balancing} condition in binary search trees.
Indeed, there exist strictly-balanced parallel bulk-update implementations, while up-to-date
concurrent algorithms only maintain \emph{relaxed} AVL condition~\cite{bouge1998height}.
Finally, we argue that complexity bounds of the data structures designed using flat parallelization
can be computed easier using the notions of \emph{work} and \emph{span},
well-established in the PRAM model.

A more thorough evaluation analysis of our approach is indispensable.
We should run our experiments on a larger scale, to check if the
performance gap with considered algorithms continues to grow. Also, we
need to enable a comparison with C++ implementations, e.g.,  the one by
Lindon and Johnsson~\cite{linden2013skiplist}, believed to be the best
concurrent \textsf{priority queue} to date~\cite{gruber2015practical}. 
Finally, we should explore the potential of the technique on
other data structures.